\documentclass[preprint,12pt,3p]{elsarticle}

\usepackage{graphicx}
\usepackage{amssymb}

\usepackage{lineno}

\usepackage{xcolor}

\usepackage{algorithm} 
\usepackage{algpseudocode} 
\usepackage{color}

\usepackage{multirow}
\usepackage{caption}
\usepackage{subcaption}
\usepackage{tabularx}

\usepackage{amsmath}
\DeclareMathOperator*{\argmax}{arg\,max} 

\usepackage{scalerel,amssymb}

\def\msquare{\mathord{\scalerel*{\Box}{gX}}}

\usepackage{hyperref}

\journal{arXiv}

\begin{document}

\begin{frontmatter}


\title{Estimating Wildfire Evacuation Decision and Departure Timing Using Large-Scale GPS Data}



\author[UF1]{Xilei Zhao\corref{cor1}}
\author[UF1]{Yiming Xu}
\author[MU]{Ruggiero Lovreglio}
\author[RMIT]{Erica Kuligowski}
\author[UC]{Daniel Nilsson}
\author[UU]{Thomas Cova}
\author[UF1]{Alex Wu}
\author[UF1]{Xiang Yan}

\cortext[cor1]{Corresponding author. Address: 1949 Stadium Rd, Gainesville, FL 32611. Phone: +01 352-294-7159. Email: xilei.zhao@essie.ufl.edu.}

\address[UF1]{Department of Civil and Coastal Engineering, University of Florida, Gainesville, FL}
\address[MU]{School of Built Environment, Massey University, Palmerston North, New Zealand}
\address[RMIT]{School of Engineering, RMIT University, Melbourne, Australia}
\address[UC]{Department of Civil and Natural Resources Engineering, University of Canterbury, Christchurch, New Zealand}
\address[UU]{Department of Geography, University of Utah, Salt Lake City, UT}

\begin{abstract}
With increased frequency and intensity due to climate change, wildfires have become a growing global concern. This creates severe challenges for fire and emergency services as well as communities in the wildland-urban interface (WUI). To reduce wildfire risk and enhance the safety of WUI communities, improving our understanding of wildfire evacuation is a pressing need. To this end, this study proposes a new methodology to analyze human behavior during wildfires by leveraging a large-scale GPS dataset. This methodology includes a home-location inference algorithm and an evacuation-behavior inference algorithm, to systematically identify different groups of wildfire evacuees (i.e., self-evacuee, shadow evacuee, evacuee under warning, and ordered evacuee). We applied the methodology to the 2019 Kincade Fire in Sonoma County, CA. We found that among all groups of evacuees, self-evacuees and shadow evacuees accounted for more than half of the evacuees during the Kincade Fire. The results also show that inside of the evacuation warning/order zones, the total evacuation compliance rate was around 46\% among all the categorized people. The findings of this study can be used by emergency managers and planners to better target public outreach campaigns, training protocols, and emergency communication strategies to prepare WUI households for future wildfire events.

\end{abstract}


\end{frontmatter}


\section{Introduction}
\label{S:1}

Wildfires are a growing threat to communities around the world \citep{boustras2017fires}. Research has shown that the intensity, frequency, and social harm of wildfires have increased in recent years, largely due to climate change \citep{liu2010trends,mccaffrey2018should,ronchi2019open,kuligowski2020modelling,zhao2021using}. Meanwhile, urban and suburban growth has led to the expansion of the wildland-urban interface (WUI), leading to an increase in the number of communities vulnerable to wildfire risks \citep{radeloff2018rapid}. As climate change accelerates and the WUI expands, the consequences of wildfires are expected to worsen. For instance, the 2020 California, Oregon, Washington Firestorms burned over five million acres and destroyed thousands of buildings, prompting evacuation orders to millions of people and causing more than two dozen fatalities \citep{westernfires}. 

To improve wildfire life safety and enhance the resilience of WUI communities, it is important to understand household behavior and movement \citep{lovreglio2019modelling}. Such knowledge can inform emergency managers to develop appropriate response measures and make effective decisions in a wildfire event, such as planning traffic management strategies, sequentially issuing evacuation orders, providing support for disadvantaged travelers, and undertaking rescues. Nevertheless, significant research gaps remain regarding the study of large-scale evacuation behavior, largely due to data limitations. To date, research on wildfire evacuation behavior has commonly relied on data collection methods such as surveys, interviews, and focus groups, e.g., \citep{kuligowski2020modelling,kuligowski2021evacuation,mccaffrey2018should}. While these studies have generated valuable insights on many aspects of household behavior during wildfires, these empirical data have limitations. For example, survey data have relatively small sample sizes (e.g., hundreds of data points), making any fitted decision models sensitive to noise and outliers. Additionally, survey data generally provide a low-resolution timeline (e.g. 2--6 hour resolution) of household decisions over the course of the evacuation \citep{lovreglio2020calibrating,fu2007modeling}. In many instances, it can be difficult or nearly impossible for some people to remember their detailed spatiotemporal trajectories on an hourly basis during an evacuation.

We aim to complement the existing studies that used surveys and mixed methods (interviews and focus groups) by leveraging an emerging data source---GPS data---that contains millions of location signals from mobile devices (e.g., smartphones and smartwatches). GPS data has shown great potential for estimating and understanding evacuation behavior for different types of disasters, e.g., hurricanes and earthquakes \citep{horanont2013large,yabe2016estimating,yabe2020effects}. However, there lacks a comprehensive and systematic methodology that is capable of using the granular spatiotemporal information of people's movements to analyze the wildfire evacuation process. 

We propose a novel methodology to apply GPS data to estimate wildfire evacuation decisions (i.e., whether to evacuate) and the corresponding departure times, where two algorithms are developed, including the home-location inference algorithm and the evacuation-behavior inference algorithm. By analyzing the movements of local residents before, during, and after the wildfire event, we categorize the evacuees into distinct groups to advance knowledge of wildfire evacuation processes. A case study of the 2019 Kincade Fire is provided to test and demonstrate the proposed methodology. This new methodology takes into account various spatiotemporal constraints to provide a comprehensive evacuee categorization, setting a foundation for future work in conducting in-depth analysis of population evacuation patterns. The results of this study can be used by emergency managers and policy makers to better understand wildfire evacuation processes for more effective evacuation planning and management.

The remaining paper is structured as follows: Section 2 reviews the related studies. Section 3 introduces the methodological framework along with two key algorithms for home-location inference and evacuation-behavior inference (for evacuation decision and departure timing), respectively. Section 4 presents the case study of the 2019 Kincade Fire, CA. Section 5 discusses the key findings of the research and concludes the paper with strengths, limitations, and future research directions.

\section{Literature Review}
The first subsection provides a brief review of existing studies of wildfire evacuation decision-making and departure timing via non-GPS means (such as surveys, interviews, and traffic counts). The second subsection of the literature review summarizes the different techniques used in prior work to estimate individual mobility patterns in day-to-day normal conditions using GPS data. The third subsection discusses the current state of research pertaining to GPS-data-based evacuation behavior analytics in emergency conditions. 

\subsection{Assessing Wildfire Evacuation Decision-Making and Departure Timing via Non-GPS Data: A Brief Review}

Existing studies of wildfire evacuation decision-making and departure timing are mainly based on non-GPS data, e.g., surveys, interviews, and traffic counts \citep{strahan2019protective,mccaffrey2018should,kuligowski2020modelling,lovreglio2020calibrating,kuligowski2021evacuation,wong2020understanding,woo2017reconstructing,grajdura2021awareness,vaiciulyte2021cross,toledo2018analysis,mclennan2019should,wong2020review}. For example, \citet{toledo2018analysis} conducted a survey study to analyze the choice whether or not to evacuate and related decisions during a wildfire event that occurred in Haifa Israel. \citet{mccaffrey2018should} surveyed homeowners in three areas in the U.S. that recently experienced a wildfire in order to understand what factors might influence people's evacuation decisions. \citet{kuligowski2020modelling} conducted a survey to assess householders' evacuation decision-making in the 2016 Chimney Tops 2 fire in Gatlinburg, TN. \citet{wong2020review} surveyed householders about their evacuation choices for three wildfires that took place in California from 2017 to 2019.

Additionally, \citet{woo2017reconstructing} applied traffic count data collected from automatic traffic recorders on highways to construct cumulative departure S-curves during the May 2016 wildfire in Fort McMurray in northern Alberta,
Canada. \citet{grajdura2021awareness} used interview and survey data to model people's awareness time, departure time, and preparation time during the 2018 Camp Fire, CA. \citet{vaiciulyte2021cross} conducted a cross-cultural comparison (between Southern France and Australia) of behavioural itinerary actions and times in wildfire evacuations using a survey approach.

The existing work has laid a solid foundation for us to understand the wildfire evacuation decision-making and departure timing. As the emerging big datasets, such as the GPS data, become available, it promises an unique opportunity to enhance our knowledge of wildfire evacuation processes by leveraging the highly granular spatiotemporal information of people's movements.

\subsection{Analyzing Non-Emergency Mobility Patterns Using GPS Data}

There are many papers that have applied GPS data to analyze and model human travel behavior. For example, \citet{calabrese2013understanding} and \citet{demissie2019understanding} applied GPS data to understand individual human mobility patterns, and, particularly, \citet{zhao2020long} used GPS data to investigate commuter trends in Beijing, China.  Regardless of the application, the techniques used to analyze mobile phone location data are quite similar. A user’s travel behavior can be broken down into two simple categories: stays and trips \citep{wang2018applying}, which are the fundamental building blocks of analyzing travel behavior with GPS data. Many research papers discuss this topic and define a “stay” as a user remaining stationary for a given time threshold while “trips” are the movements between two stays \citep{wang2018applying, zhao2020long,demissie2019understanding,chen2016promises}. These “stays” are geographic locations with which the user interacts and there are several techniques used in research to extract locations of importance such as home location, work location, and shopping locations. The following paragraphs will briefly describe the most popular techniques used to model stays and trips.

\subsubsection{Clustering}
Researchers use clustering to group GPS data both by space and time \citep{vanhoof2018assessing, wang2018applying, yabe2019cross, xu2015understanding, chen2016promises, tettamanti2012route, wang2010transportation,ahas2010using}. For example, a home-location inference algorithm may infer home location by grouping areas of frequent return at night for multiple days in a row \citep{vanhoof2018assessing, yabe2019cross}. In a similar fashion, a work location may be determined using a clustering algorithm that analyzes a user’s weekday GPS data and detects the most frequently visited location \citep{xu2015understanding, wang2010transportation, chen2016promises}. To take this one step further, the clustering algorithm can be overlayed with land use data in order to detect daytime locations other than work such as schools and restaurants \citep{chen2016promises, alexander2015origin}. 

\subsubsection{Time-Space Heuristics}
A common approach to detect home locations using GPS data is to use simple rule-based algorithms (also called time-space heuristics). These simple rules are often applied in conjunction with clustering to determine the type of location detected \citep{vanhoof2018assessing, wang2018applying, zhao2020long, yu2020mobile, yabe2019cross, demissie2019understanding, xu2015understanding}. For example, home location can be detected by observing where the greatest number of GPS signals occur during hours of the night, more specifically, a time threshold such as from 12 am to 4 am in which the user is most likely to be home \citep{yu2020mobile, li2014framework}. 

However, a major limitation is that rule-based algorithms are generalizations that introduce bias into the study \citep{vanhoof2018assessing}. For example, if a home detection algorithm examines where users spend most of their time during the night, this rule would not be accurate for people who have night jobs and rest during the day \citep{wang2018applying}. With that being said, the rule-based algorithms are generally accepted in this field of study for two main reasons: ease of implementation and limited validation techniques make it difficult to evaluate the accuracy of more complex models \citep{vanhoof2018assessing}. 

\subsubsection{Map Matching}
The most common way to determine a trip route is called “map matching.” This method matches the progression of GPS location nodes with a line that follows the most logical nearby roads. The more nodes present, the more accurate the route \citep{wang2018applying}. Additionally, the combination of the approximate speed of the user, the surrounding infrastructure along the trip route, and the overall geographic location (e.g., water, urban, rural) can be used to detect the travel mode \citep{quddus2007current}.

\subsection{Modeling Evacuation Behavior Using GPS Data}

Similar to using mobile phone location data (GPS data) for travel behavior analysis in normal conditions, GPS data also has great potential for evacuation studies through (1) real-time evacuation monitoring and (2) using historical GPS data to investigate evacuation behavior during previous emergencies. Although there are examples of analyzing general mobility patterns using GPS data under normal conditions, there is limited research on the application of GPS data to emergency evacuation during disasters and wildfires in particular.

Through the last decade, researchers have started using historical GPS data for investigating emergency evacuation. \citet{hayano2013estimation} used GPS data to measure the total number of people moving in and out of the evacuation zone during the Fukushima Nuclear Power Plant Accident, \citet{yabe2019cross} used mobile phone location data to analyze evacuation behavior after earthquakes, \citet{yabe2020effects} used more than 1.7 million mobile phone's GPS data to investigate the effect of income inequality on evacuation behavior during Hurricane Irma, and \citet{song2013intelligent} used GPS data of 1.6 million users to analyze and simulate evacuations during the Great East Japan Earthquake and the Fukushima Daiichi nuclear accident. \citet{horanont2013large} and \citet{yabe2016estimating} also investigated the benefits and how GPS data could be leveraged to analyze evacuation behavior in real-time. Real-time information can give decision-makers the insight needed to determine where to spend more of their efforts during an emergency. 

However, little research has been focused on creating a comprehensive methodology that can systematically evaluate wildfire evacuation processes and extract insights regarding different types of evacuees (e.g., evacuees who left home before the official warning/order and evacuees who lived outside of any evacuation warning/order zones).

\section{Methodology}
In this section, we first present the overall methodological framework for estimating wildfire evacuation decisions (whether to evacuate) and departure times using GPS data in Section 3.1. We then discuss the home-location inference algorithm and the evacuation-behavior inference algorithm in Sections 3.2 and 3.3, respectively. 

\subsection{Methodological Framework}

\begin{figure}[H]
    \centering
    \includegraphics[width=0.6\textwidth]{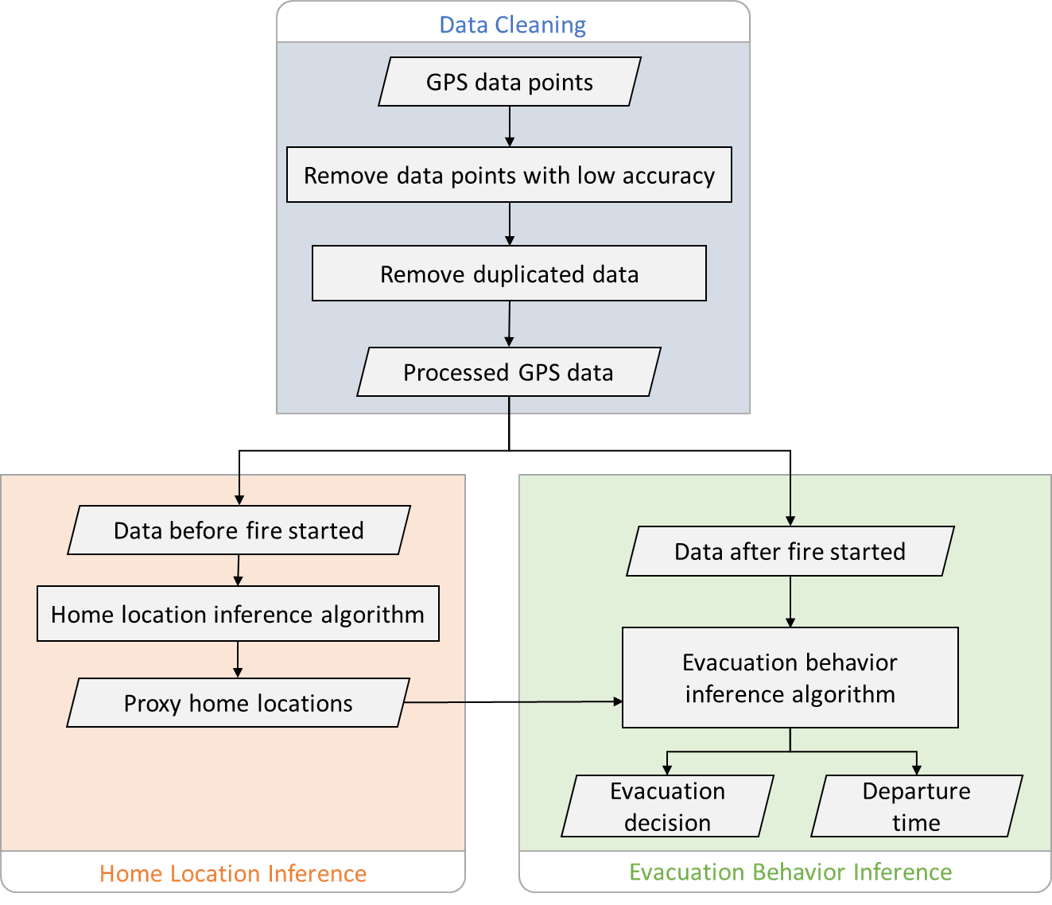}
    \caption{Overall methodological framework for estimating wildfire evacuation decision and departure timing}
    \label{fig:frw}
\end{figure}

The first major step of the proposed methodological framework is data cleaning (blue box in Figure \ref{fig:frw}). More specifically, we first remove the inaccurate data points. As GPS records usually have spatial measurement errors \citep{zhang2016spatial}, some data providers label the accuracy of the latitude and longitude of a GPS record, measured by distance error. Then, modelers can choose a distance error threshold to filter out highly inaccurate data points. Note that this step can be skipped if the data provider does not provide this data field. However, skipping this the data cleaning process might compromise the reliability of the analysis carried out on the data to investigate evacuees’ behavior. After removing inaccurate data points, we remove the duplicated records according to device ID, timestamp, and location.  

After the data cleaning process, we divide the processed dataset into two subsets: the records before the fire started and the records after the fire started. The data before the start of the fire is used to infer residents' proxy home locations, where we develop a home-location inference algorithm. The inferred home locations and the data after the fire started are used as the inputs to infer individual-level evacuation behavior. In this work, we propose a novel methodology to estimate residents' evacuation decisions and the corresponding departure times of the evacuees. The home-location inference algorithm is explained in Subsection 3.2, and the evacuation-behavior inference algorithm is described in Subsection 3.3.

\subsection{Home-Location Inference}

As discussed in the literature review section (Section 2.1), there are two common approaches to infer home locations using GPS data. The first approach uses clustering in combination with supplemental information such as land use data to infer regular activities and the resident's home location \citep{calabrese2013understanding,wang2018applying}. This approach is computationally complex because it identifies all common activity locations (e.g., home location, work location, etc.). The second approach is called time-space heuristics \citep{xu2015understanding,ahas2010using}. This rule-based method is commonly used to detect home locations and is often applied in conjunction with clustering \citep{yu2020mobile,li2014framework}. In this study, we adopt the second approach to infer home locations using the time-space heuristics method accompanied by clustering.

\begin{figure}[H]
    \centering
    \includegraphics[width=0.85\textwidth]{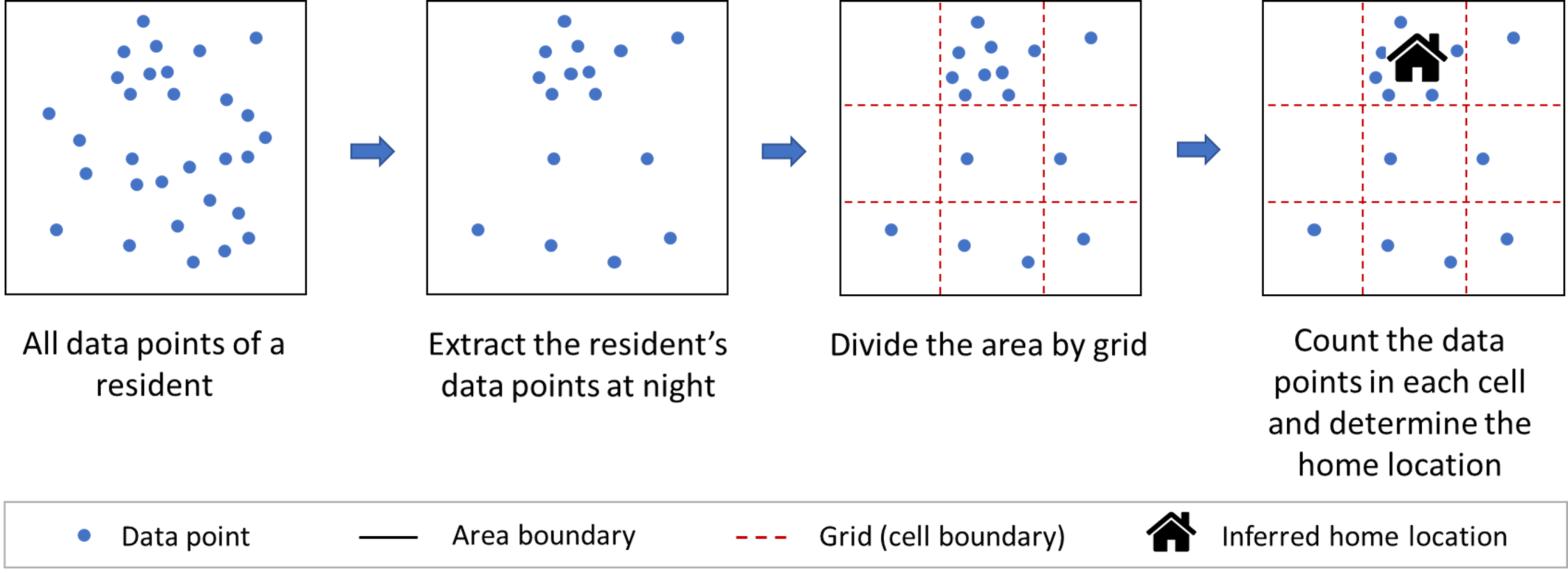}
    \caption{Home-location inference algorithm}
    \label{fig:algo_home}
\end{figure}

The process of determining the resident's home location adopted in this study is presented in Figure \ref{fig:algo_home}. The home-location inference algorithm assumes that residents in this area spend most of their nighttime at home before the start of the fire. This means that the most visited place during night hours based on the resident's GPS traces becomes their predicted home location. To achieve this outcome, we first extract data points for each resident before the start of the fire. Next, we extract the resident’s data points during the nighttime (i.e., 10 pm to 6 am \citep{yu2020mobile,li2014framework}). After that, the study area is divided into square cells by a grid. The size of the cells is set to be 20 × 20 meter according to the typical size of a single-family home in the U.S. The most visited cell is defined as the cell containing the most number of data points. The centroid of the most visited cell is identified as the home location of this resident. Let $C_i$ be the location of the centroid of cell $i$. Let $N_i^j$ be the number of reported data points of resident $j$ within a given cell $i$ during the nighttime. The inferred home location $H_j$ of the resident $j$ can be defined as:

\begin{equation}
H_j=C_{p_j}
\end{equation}
\begin{equation}
p_j=\argmax_{x \in \{1,2,...,m\}} N_x^j
\end{equation}

\noindent where $p_j$ is the cell with the most data points for resident $j$ and $m$ is the number of cells in the study area.

\subsection{Evacuation-Behavior Inference}
In this study, we develop a rule-based algorithm to infer evacuation behavior of residents based on the GPS data. We will use \textit{evacuation zone} to represent the geographic area under evacuation warning/order. Note that we only analyze the evacuation behavior of people who resided in or near the evacuation zones (within 5 miles of the evacuation zones' boundaries) based on the GPS data gathered prior to the event.

The evacuation-behavior inference algorithm is based on the following assumptions: 
\\

\noindent \textit{\textbf{Assumption 1}}: All evacuees departed from home.

\noindent \textit{\textbf{Assumption 2}}: If the distance between the resident's current location and the resident’s home location was larger than $D$ (i.e., home buffer radius), the resident has left home.

\noindent \textit{\textbf{Assumption 3}}: A resident is considered as an evacuee, if they left the evacuation zone during the evacuation process.

\noindent \textit{\textbf{Assumption 4}}: The evacuation departure time is when the evacuee left home to evacuate.
\\

We divide the evacuees into four groups based on their proxy home location and evacuation departure time using the following four definitions:
\\

\noindent \textit{\textbf{Self-evacuee}}: The evacuee, located in or near the evacuation zone, left after the fire started but before any evacuation warning/order was issued.

\noindent \textit{\textbf{Shadow evacuee}}: The evacuee, located outside but near the evacuation zone, left after an evacuation warning/order was issued.

\noindent \textit{\textbf{Evacuee under warning\footnote{This proposed algorithm is based on California's standard statewide evacuation terminology and policy: \url{http://calalerts.org/evacuations.html}. Any advice whereby the warning signifies a \textit{potential} threat to life and/or property and those who require additional time to evacuate should do so and evacuation order signifies an immediate threat to life and in some cases, the lawful order to leave now (see \url{http://calalerts.org/evacuations.html}).}}}: The evacuee was in the evacuation warning zone and evacuated after the warning was issued and before an order was issued (if any).

\noindent \textit{\textbf{Ordered evacuee}}: The evacuee lived in the evacuation order zone and evacuated after the order was issued.
\\

Note that the \textit{shadow evacuee} concept is borrowed from the nuclear \citep{zeigler1981evacuation}, hazmat \citep{mitchell2007improving}, and hurricane evacuation literature \citep{Gladwin1997}, and we use it here to help us better understand the wildfire evacuation process. In addition to the evacuee categories, we also have two other resident categories, defined as follows:
\\

\noindent \textit{\textbf{Non-evacuee}}: Resident who did not evacuate, regardless of home location; i.e., inside or outside of the evacuation zone.

\noindent \textit{\textbf{Uncategorized person}}: All cases that do not fit the prior ones\footnote{For example, resident who left home after evacuation warning/order was lifted, resident who returned home before evacuation warning/order was lifted, and resident who did not have any GPS signals after a potential evacuation}.
\\

\begin{figure}[H]
    \centering
    \includegraphics[width=0.70\textwidth]{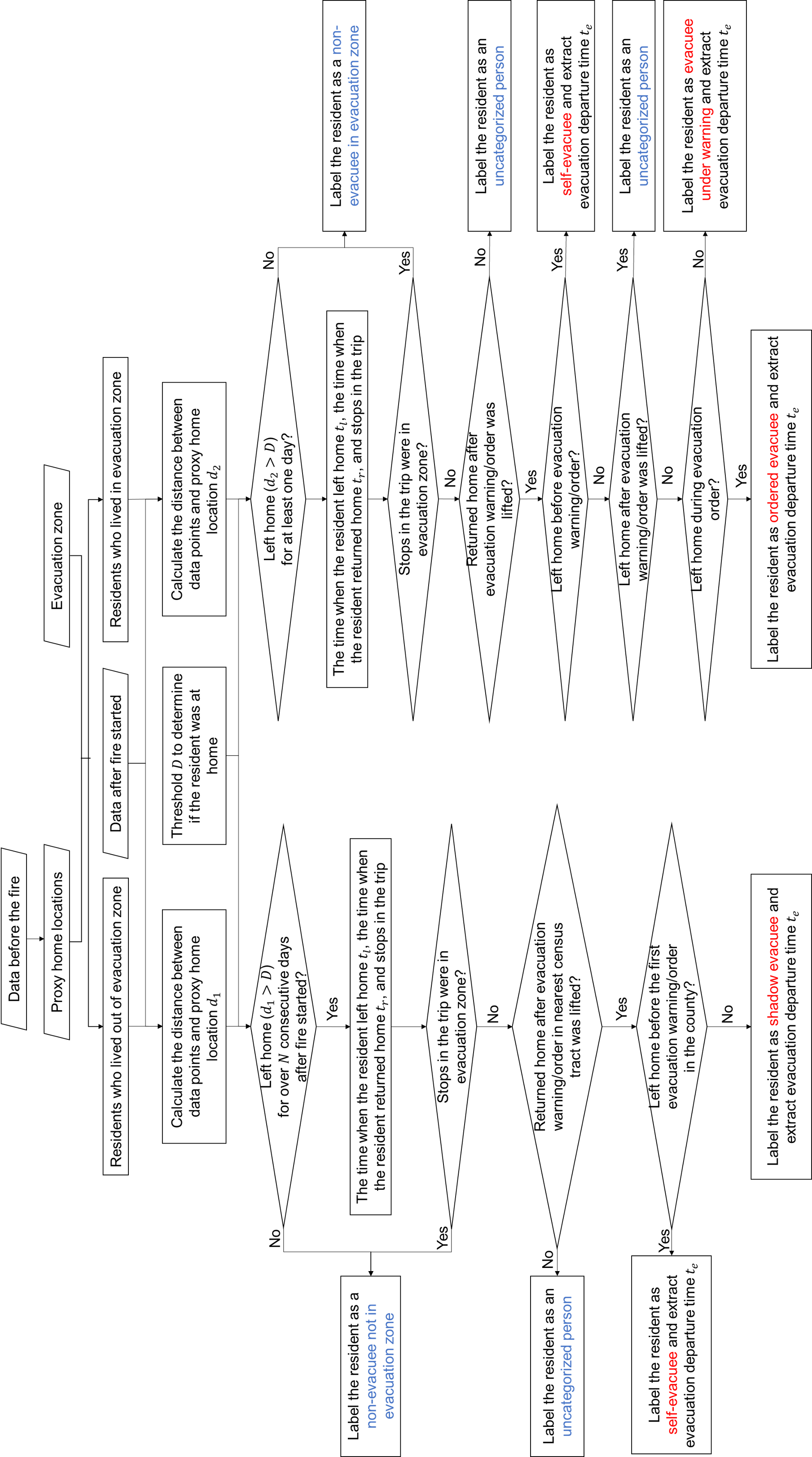}
    \caption{Evacuation-Behavior Inference Algorithm}
    \label{fig:algo_eva}
\end{figure}

\indent Given these definitions, we develop an algorithm to infer the evacuation behavior of the residents and categorize the evacuees. The process of the evacuation-behavior inference algorithm is presented in Figure \ref{fig:algo_eva}. Based on the proxy home locations and the evacuation zones, we first divide the residents into two group: \textbf{residents who lived outside the evacuation zone}, and \textbf{residents who lived in the evacuation zone}. 

For \textit{residents who lived outside the evacuation zone}, we calculate the distance between data points and the resident’s home $d_1$ and detect whether the resident ever left home for over $N$ consecutive days using the threshold $D$. If not, we label the resident as a \textit{non-evacuee not in evacuation zone}. Otherwise, we extract the time when the resident left home $t_l$ and the stops in the trip (i.e., the places where the resident stayed at night in the trip). If the stops in the trip were in the evacuation zone, we label the resident as a \textit{non-evacuee not in evacuation zone}.
Then, we compare the time when the resident return home $t_r$ to the time when the evacuation warning/order in the nearest census tract was lifted. If the resident returned home before the evacuation warning/order in the nearest census tract was lifted, we label the resident as \textit{uncategorized person}. After that, we compare the time when the resident left home $t_l$ to the time when the first evacuation warning/order in the county was issued. If the resident left home before the first evacuation warning or the order was issued in the county, we label the resident as \textit{self-evacuee}; otherwise, we label the resident as \textit{shadow evacuee}. After this, we extract the evacuation departure time $t_e$ of the resident. 

For \textit{residents who lived in the evacuation zone}, we first calculate the distance between data points and the resident’s home $d_2$ and detect whether the resident had ever left home for at least one day using the threshold $D$. If not, we label the resident as a \textit{non-evacuee in evacuation zone}. Then, we extract the time when the resident left home $t_l$, the time when the resident returned home $t_r$, and the stops in the trip. If the stops in the trip were in evacuation zone, we label the resident as a \textit{non-evacuee in evacuation zone}. If the resident returned home before the evacuation warning/order was lifted, we label the resident as an \textit{uncategorized person}. If the resident left home before the evacuation warning/order, we label the resident as \textit{self-evacuee}. If the resident left home after the evacuation warning/order was lifted, we label the resident as an \textit{uncategorized person}. If the resident left home during the evacuation warning, we label the resident as \textit{evacuee under warning} and extract the evacuation departure time $t_e$. If the  left home during the evacuation order, we label the  as \textit{ordered evacuee} and extract the corresponding evacuation departure time $t_e$. 

Based on the evacuation-behavior inference results, we can further calculate the evacuation compliance rate for each census tract. The evacuation compliance rate $\alpha_t^i$ on a given time period $t$ in a given geographical area $i$ can be calculated by:

\begin{equation}
\alpha_t^i=\frac{M_t^i}{N_t^i}
\end{equation}

\noindent where $M_t^i$ is the number of evacuees who left during time period $t$ in area $i$, $N_t^i$ is the total number of residents living in area $i$ during time period $t$.

\begin{figure}[H]
    \centering
    \includegraphics[width=0.75\textwidth]{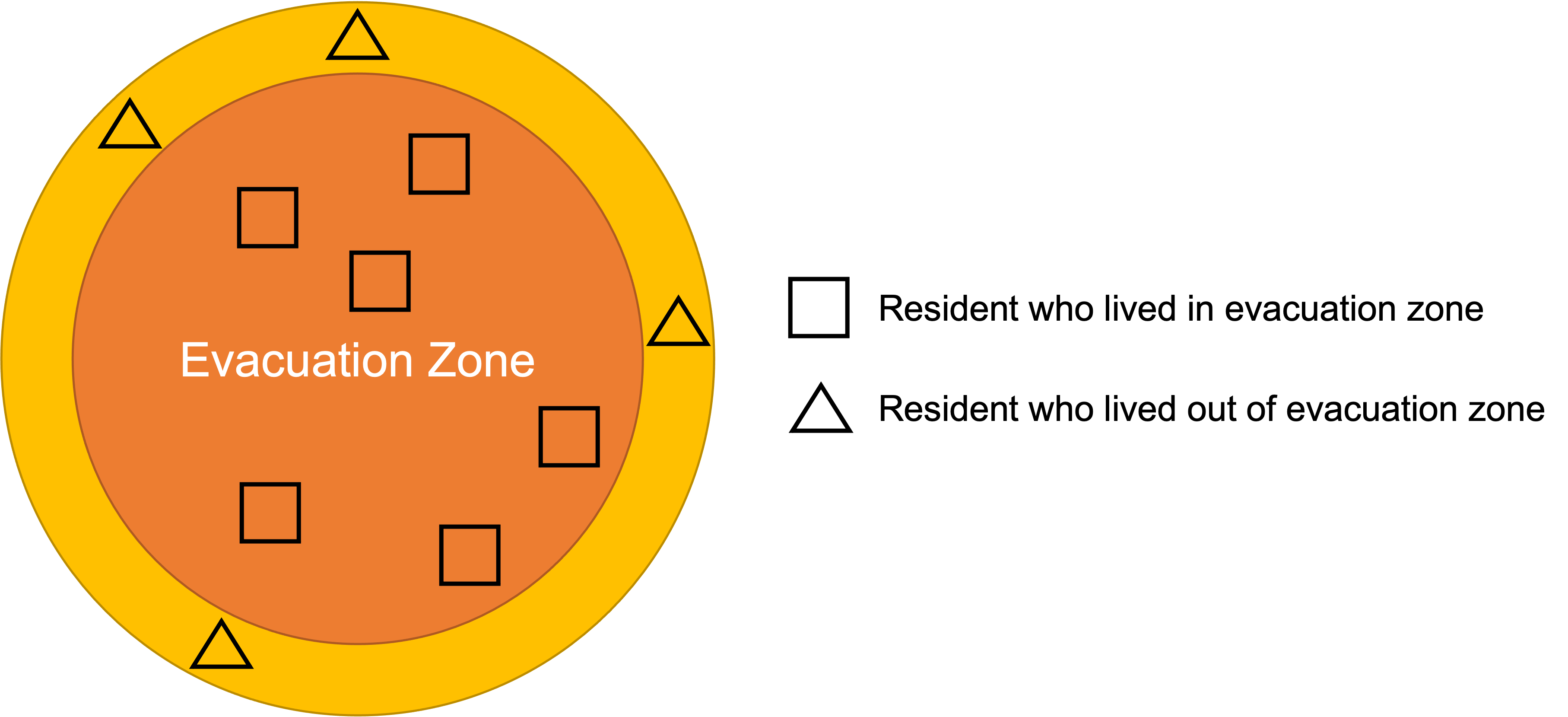}
    \caption{An Example to Illustrate the Categorization of Evacuee Groups}
    \label{fig:example}
\end{figure}

Here, we use a simple, hypothetical example to explain how we categorize these four evacuee groups. As shown in Figure \ref{fig:example}, there is an evacuation zone in orange with a five-mile buffer in yellow. We use squares to indicate s who lived in the evacuation zone, and triangles to denote s who lived outside but near the evacuation zone. The fire started on Day 0, the warning was issued on Day 3, and the evacuation order was declared on Day 5. The definitions of different evacuee groups based on the spatiotemporal constraints are shown in Table \ref{tab:eva_groups}, provided all the other constraints of evacuees are satisfied. For people who lived in the evacuation zone, they could be a self-evacuee (if they leave after the ignition of fire but before the issuance of a warning for the evacuation zone), evacuee under warning (if they leave after the warning and prior to the order for the evacuation zone), or ordered evacuee (if they leave after the order for the evacuation zone). For people who lived outside the evacuation zone, they were either self-evacuees or shadow evacuees, depending on the timing of the first warning in the entire impacted area. 

\begin{table}[H]
\centering
\caption{Definitions of Different Evacuee Groups}
\begin{tabular}{|l|l|l|l|l|}
\hline
            & \textbf{Day 1} & \textbf{Day 2} & \textbf{Day 3: Warning}        & \textbf{Day 5: Evacuation order}  \\ \hline
$\msquare$  & Self-evacuee   & Self-evacuee   & Evacuee under warning & Ordered evacuee \\ \hline
$\triangle$ & Self-evacuee   & Self-evacuee   & Shadow evacuee        & Shadow evacuee  \\ \hline
\end{tabular}
\label{tab:eva_groups}
\end{table}

\section{Case Study and Results}

This section provides an overview of the case study used in this study (i.e., the 2019 Kincade Fire) in Section 4.1. The GPS data used to investigate this fire is described in Section 4.2 while the results regarding the home location and the evacuation estimations are provided in Section 4.3 and Section 4.4.

\subsection{Study Site Exploration}
\begin{figure}[H]
    \centering
    \includegraphics[width=0.65\textwidth]{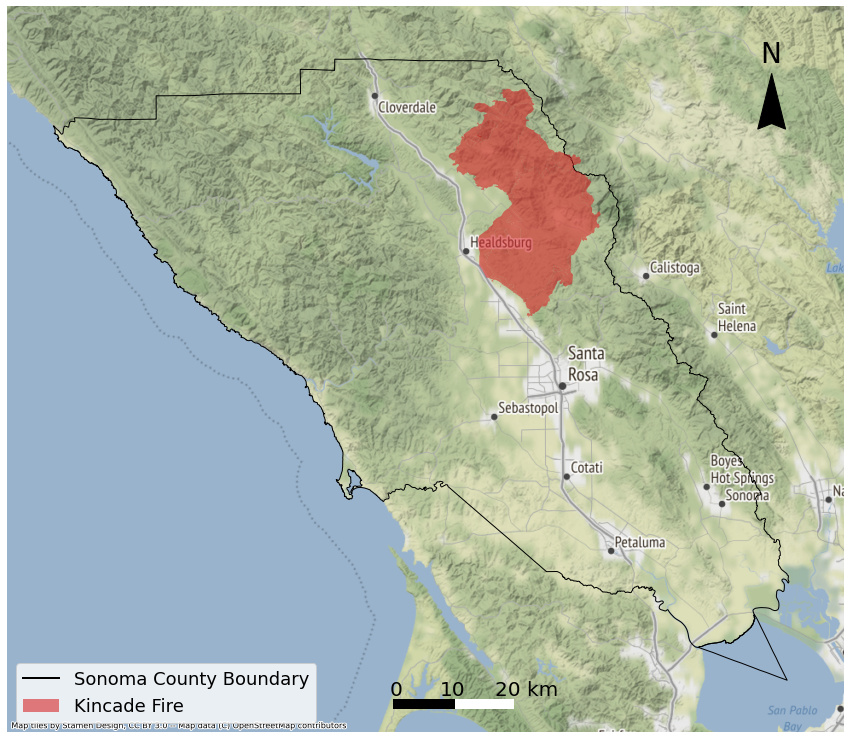}
    \caption{Sonoma County and the Kincade Fire Perimeter}
    \label{fig:sonoma}
\end{figure}

We selected the 2019 Kincade Fire, Sonoma County, CA, as the case study. Sonoma County is located in Northern California, U.S. According to the U.S. Census Bureau, Sonoma County's population estimate in 2019 was 494,336, and its county seat and largest city is Santa Rosa. The highway system of Sonoma County consists of U.S. Highway 101, and State Highways 1, 12, 37, 116, 121, and 128. The Kincade Fire started northeast of Geyserville at 9:27 p.m. on October 23, 2019 and was fully contained at 7:00 p.m. on November 6, 2019. The fire burned 77,758 acres, destroyed 374 structures, damaged 60 structures, and caused 4 injuries \citep{sonoma2020report}. As the fire spread, the mandatory evacuation order was first issued in Geyserville on October 26, and then the evacuation warnings and orders grew to encompass nearly all of Sonoma County in the following days, making it the largest evacuation in Sonoma County's history. The study site and the fire perimeter are shown in Figure \ref{fig:sonoma}.

\subsection{Data Description and Cleaning}
The GPS data\footnote{The GPS data underwent Gravy's cleansing processes and was optimized with Gravy Location Data Forensics--filtering and categorizing inaccurate and even fraudulent location signals. This enabled us to identify and use only the cleansed location signals relevant to this project.} was provided by Gravy Analytics and built on privacy-friendly mobile location data. Gravy’s location data platform processes raw location signals from multiple data providers representing over 150 million U.S. mobile devices. 
After the data cleaning process (i.e., removing the data points with errors greater than 250 meters and duplicated observations), we included 100,913,550 GPS signal records in Sonoma County, CA from October 16, 2019 to November 13, 2019 for analysis. The fields of the GPS data include the unique identifiers for devices, latitude, longitude, the geohash (a geocode format\footnote{More details about geohash can be found here: \url{http://geohash.org/site/tips.html}.} using a short alphanumeric string to express a location), timestamp, time zone, and Forensic Flag (which indicates the accuracy of location signals).

\begin{figure}[H]
    \centering
    \includegraphics[width=0.80\textwidth]{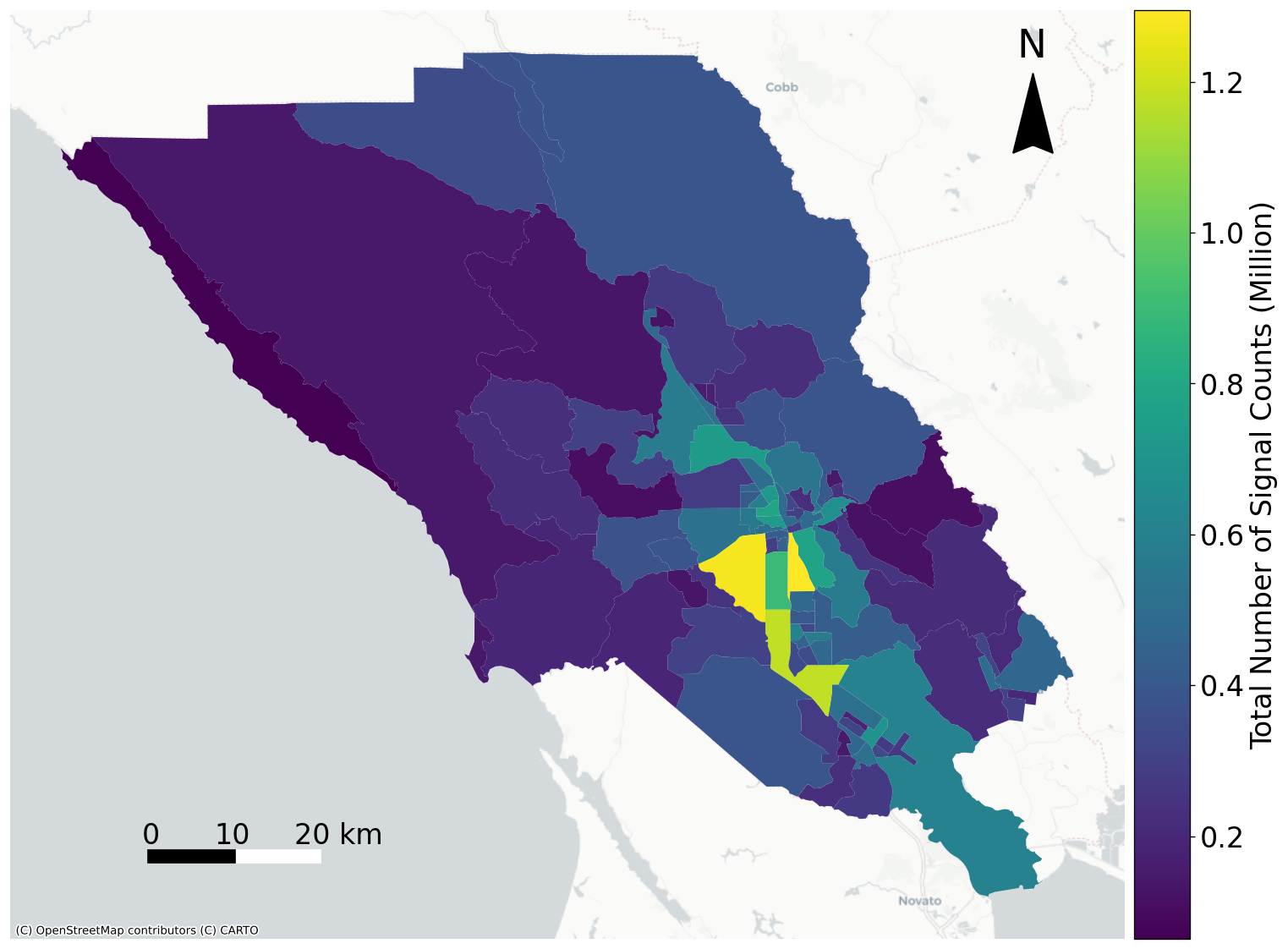}
    \caption{Distribution of Total Signal Counts for Residents at the Census Tract Level in Sonoma County, CA}
    \label{fig:signal}
\end{figure}

To ensure the reliability of the inference, we only used the records of daily frequent users of mobile devices in this study. A daily frequent user is defined as a user who had at least 20 signals on each day before the fire (i.e., from 10/16/2019 to 10/23/2019). These users are considered as local residents in this study. After this step, we retained 44,211,050 records, or a total of 5,338 residents. The distribution of these data points is shown in Figure \ref{fig:signal}. It shows a higher number of total signal counts in census tracts with higher population densities. 

\subsection{Home-Location Inference}
\begin{figure}[H]
    \centering
    \includegraphics[width=0.65\textwidth]{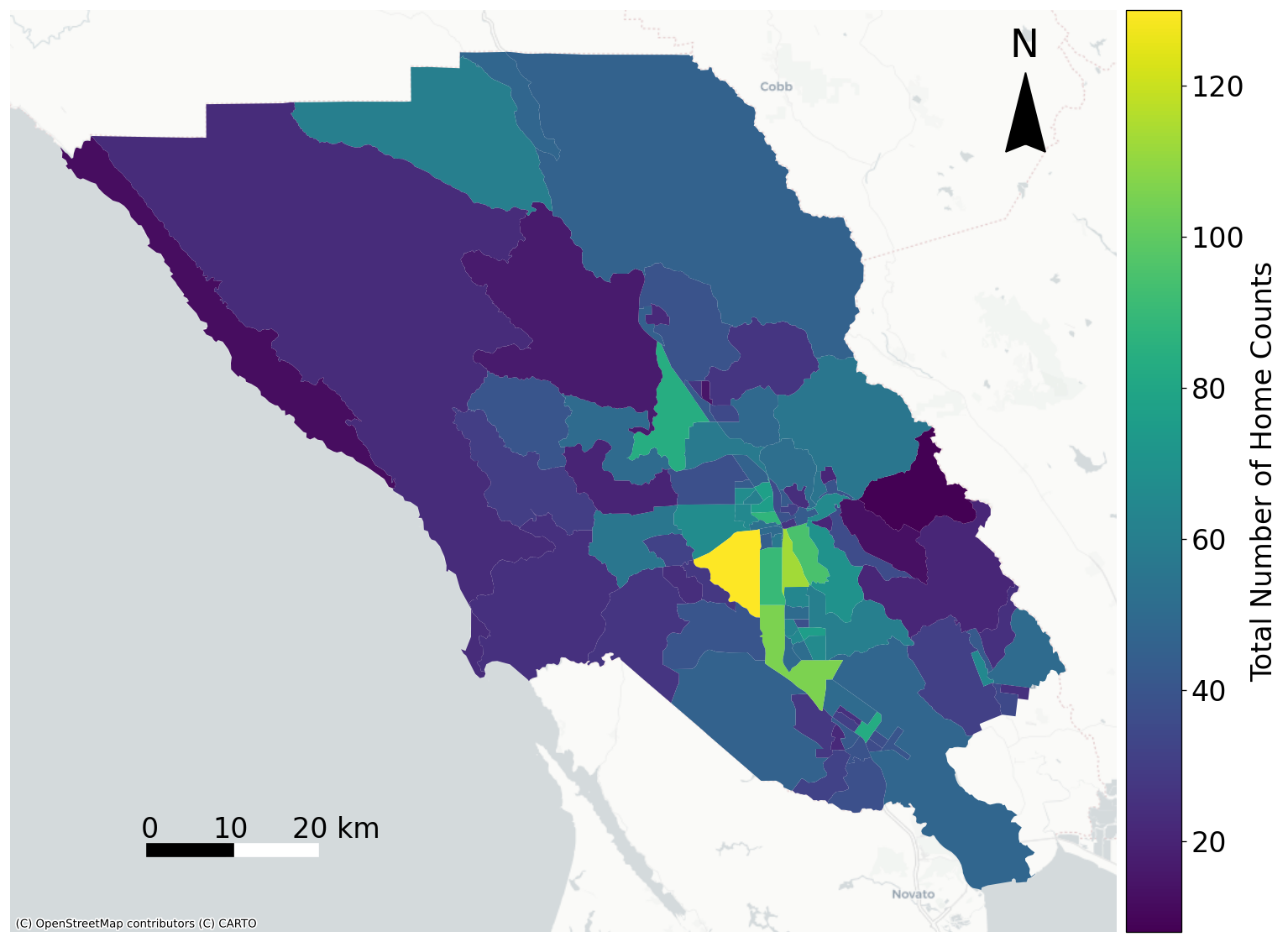}
    \caption{Distribution of Proxy Home Locations at the Census Tract Level}
    \label{fig:home_loc}
\end{figure}

By applying the home-location inference algorithm proposed in Section 3.2, we estimated the proxy home locations of the residents (i.e., daily frequent users) in Sonoma County, CA. Figure \ref{fig:home_loc} illustrates their distribution at the census tract level. We identified a total of 5,166 homes/residents in Sonoma County, accounting for 1.05\% of the total county population in 2019 \citep{census2019}. Note that for different residents who lived in the same household, we double-counted the same home location in Figure \ref{fig:home_loc}. We had home location observations in all the census tracts of Sonoma County, CA, but 5\% census tracts had less than 20 inferred homes, making the following analyses of these tracts less reliable due to the uncertainties within the sample.  

\begin{figure}[H]
    \centering
    \includegraphics[width=0.65\textwidth]{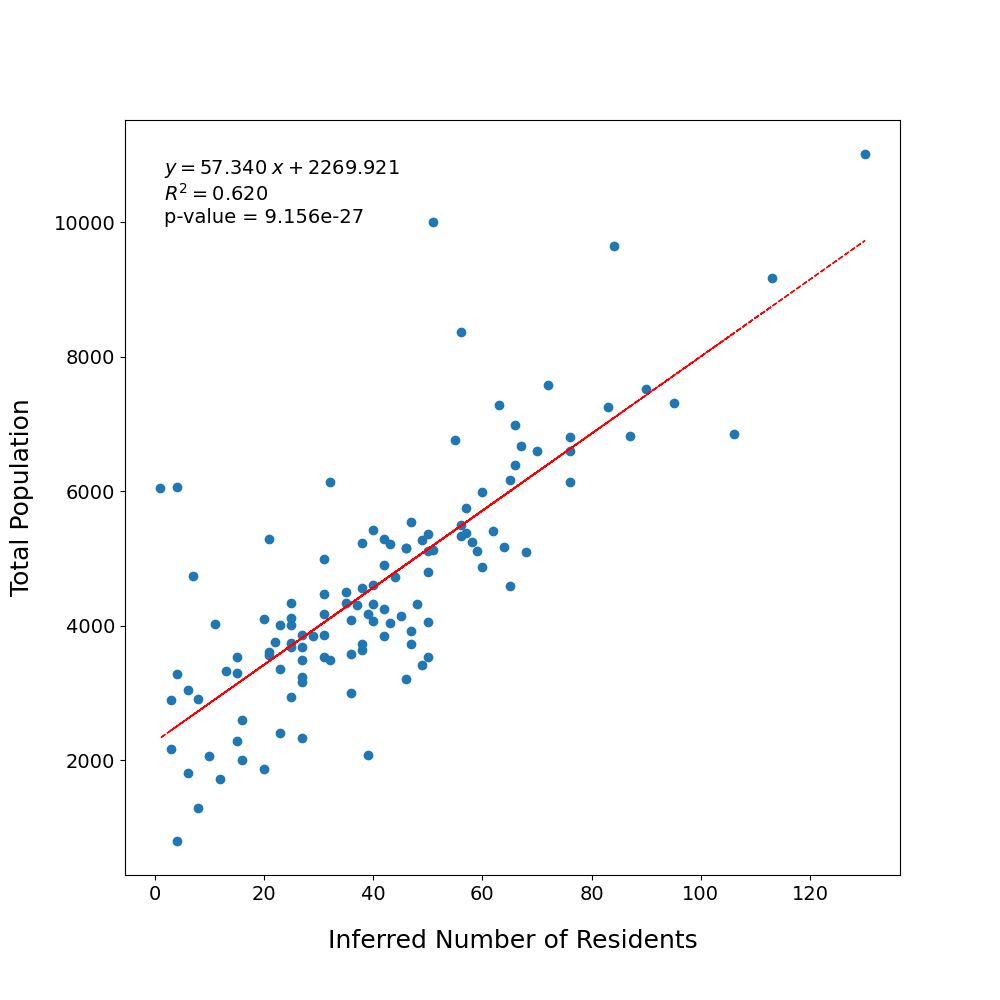}
    \caption{Relationship Between Inferred Number of Residents Versus Total Population at the Census Tract Level in Sonoma County, CA}
    \label{fig:sam1}
\end{figure}

\begin{figure}[H]
    \centering
    \includegraphics[width=0.65\textwidth]{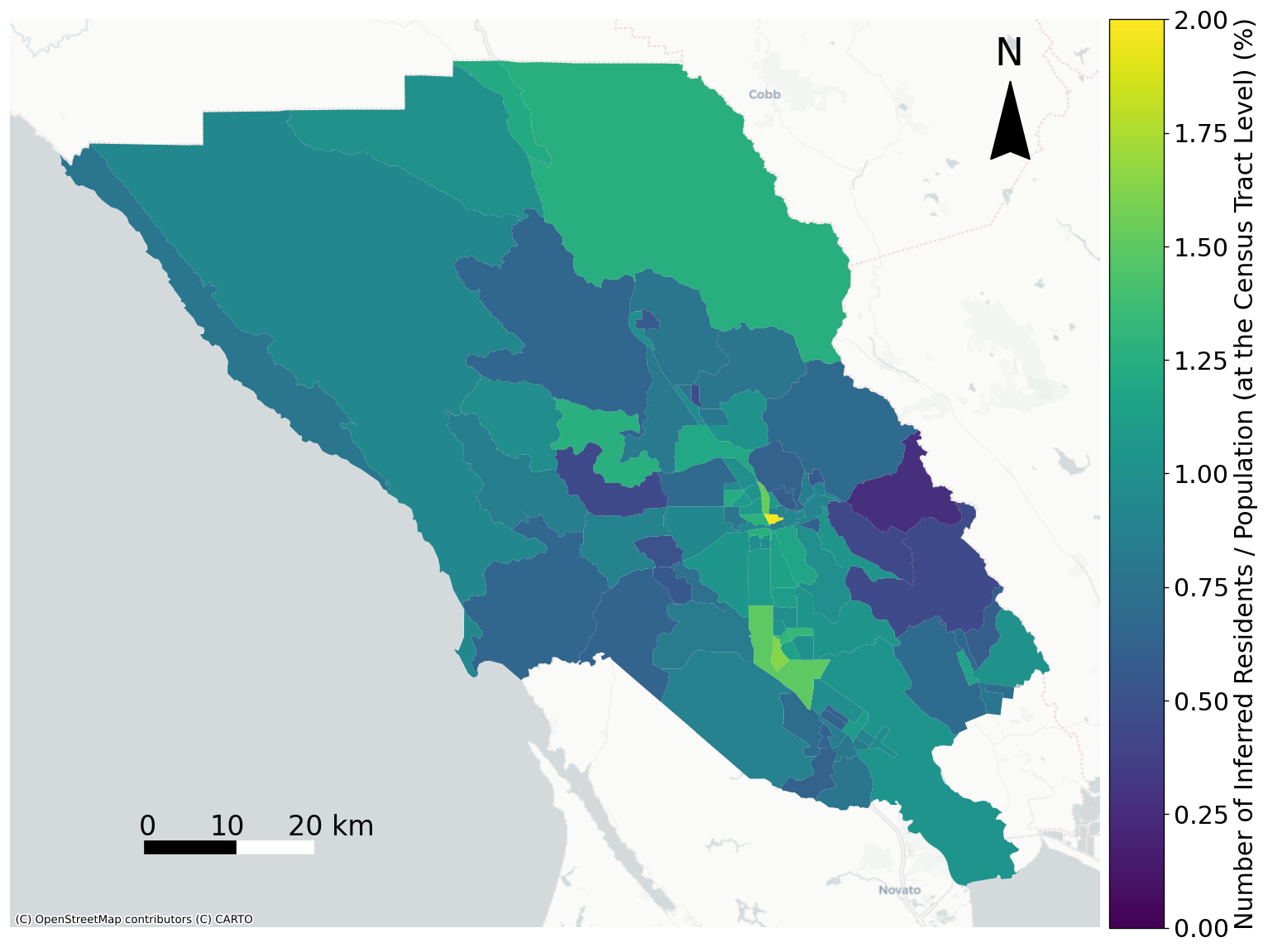}
    \caption{Distribution of the GPS Data Sampling Rates at the Census Tract Level}
    \label{fig:sam2}
\end{figure}

To examine the sampling bias of the data, we fitted a simple linear regression model between the inferred number of residents (equal to the proxy home locations) and the total population at the census tract level (see Figure \ref{fig:sam1}). The $R^2$ of the linear regression model is 0.620\footnote{The $R^2$ will equal to 1, if there is no sampling bias.} and the $p$-value of the beta coefficient is extremely small (9.156e-27)\footnote{For $p$-value less than 0.001, we can conclude the beta coefficient is statistically significant.}, which suggests relatively low sampling bias of the GPS data. However, we observed more outliers above the fitted line in Figure \ref{fig:sam1}, indicating some census tracts had a smaller number of inferred residents compared to their total population (i.e., low GPS data sampling rates). We further estimated the census-tract-level sampling rates and displayed them in Figure \ref{fig:sam2}. It is clear that most low-sampling-rate areas were located around Santa Rosa, especially in the southeast. 


\subsection{Evacuation Estimation}
Based on Assumptions 1-5 and the evacuation-behavior inference algorithm illustrated in Figure \ref{fig:algo_eva}, we identified different groups of evacuees and their corresponding departure times. We set the two main parameters of the algorithm as follows: $N = 2$ days (the time threshold for residents who lived outside evacuation zone to determine how long the resident left home), and $D = 5$ miles or 8 km (the distance threshold to determine how far the resident was away from home). More specifically, according to Figure 4 in \citet{wong2020review}, the majority of evacuees had to drive more than 1 hour during their evacuations in the 2017 Northern California Wildfires, the 2017 Southern California Wildfires, and the 2018 Carr Wildfire; we thus assumed $N = 2$ days. 

\subsubsection{Temporal Patterns}

\begin{figure}[H]
    \centering
    \includegraphics[width=1\textwidth]{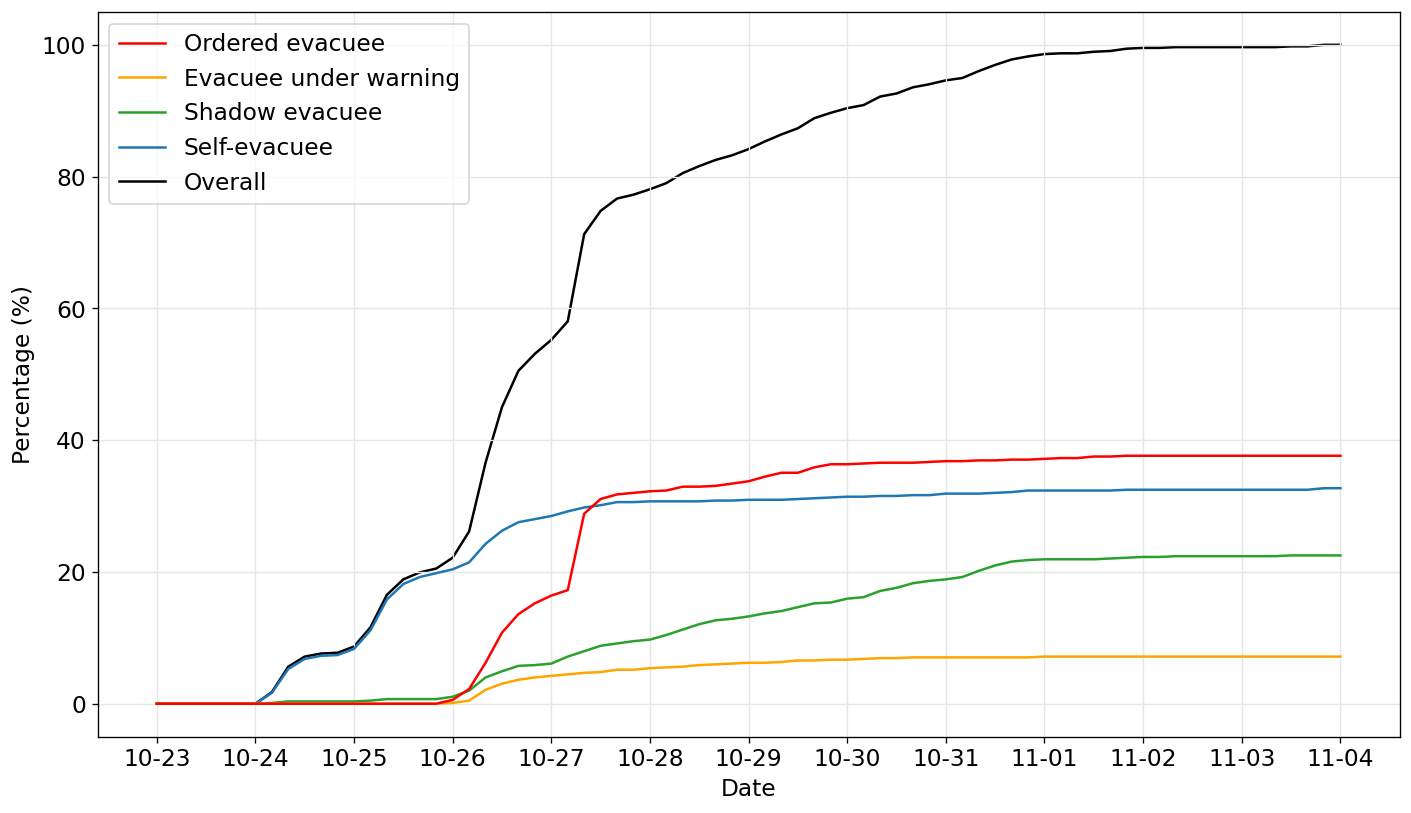}
    \caption{12-Day Cumulative Evacuation Response Curves of Overall Evacuees and Four Different Evacuee Groups}
    \label{fig:temporal}
\end{figure}

The response time of the householders' evacuation is traditionally estimated using cumulative departure S-curves (e.g., the Rayleigh distribution), which are based on empirical data collected during different hurricanes \citep{murray2013evacuation,ozbay2012use}. Cumulative S-curves have been also applied for wildfire evacuations in multiple studies \citep{woo2017reconstructing,cova2011modeling,dennison2007wuivac,wolshon2007emergency}. However, there are several drawbacks to this approach \citep{fu2004development,yazici2008evacuation}.
For instance, S‐curves were originally created to capture the evacuation departure timing within a day \citep{ozbay2012use,wolshon2007emergency}, making them unsuitable for the staged evacuation process which might take days. Interestingly, \citet{dixit2011validation} analyzed the traffic count data from Southeast Louisiana observed during the Hurricane Katrina evacuation and showed back-to-back (or, double) S-curves to represent the cumulative evacuation response over a 2-day period. Therefore, in this study, we generate 12-day cumulative evacuation response curves (for overall evacuees as well as different evacuee groups) to capture the entire wildfire evacuation process, as illustrated Figure \ref{fig:temporal}.

According to Figure \ref{fig:temporal}, the overall response curve (black curve) is an aggregation of multiple S-curves, where the S-curve of October 26 has the largest slope. From our sample, we found that people started to self-evacuate on October 24, 2019 as soon as they heard of the fire, which was ignited in the evening of October 23, 2019. The total number of self-evacuees stabilized after October 27, 2019. Additionally, we found that shadow evacuees started to emerge on October 26, 2019 and gradually grew to the maximum on October 31, 2019. Self-evacuees (blue curve, 33\%) and shadow evacuees (green curve, 23\%) accounted for more than half of the total evacuees (55\%). The large numbers of self-evacuees and shadow evacuees suggest that the local residents were sensitive to wildfire risks due to prior wildfire experience (i.e., the 2017 Tubbs Fire) \citep{erica2021}. We also observed that a non-trivial amount (7\%) of evacuees left home as soon as they received the evacuation warnings (yellow curve). This finding has been corroborated by the local emergency management officials, as their warning message was to recommend that people who needed extra time to leave should evacuate once they received the evacuation warnings. Most ordered evacuees left home on October 26 and 27, 2021 once they received the mandatory evacuation orders (red curve), and this evacuee group accounted for 38\% of the total evacuees. 

\subsubsection{Spatial Patterns}

\begin{figure}[H]
    \centering
    \includegraphics[width=0.75\textwidth]{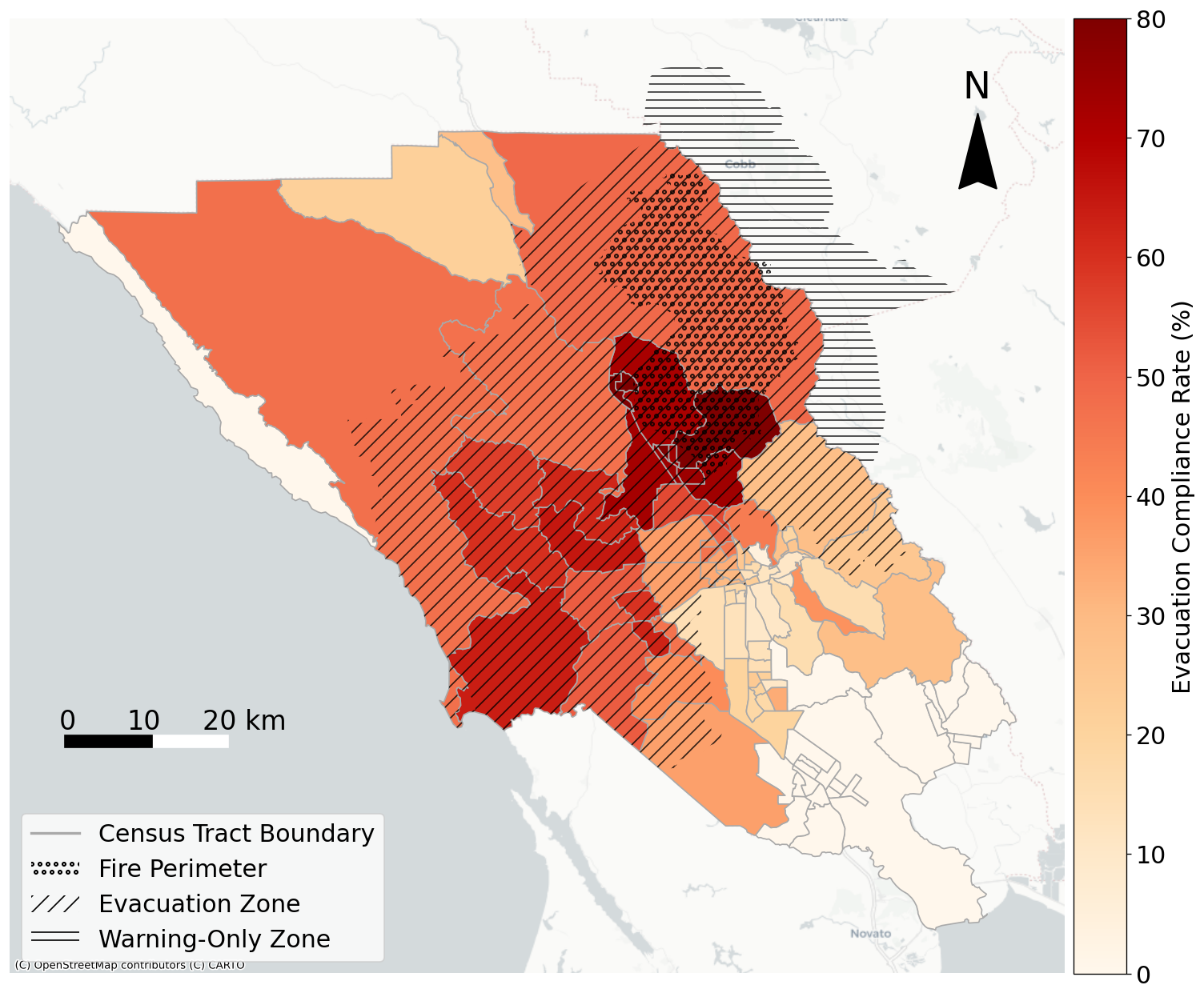}
    \caption{Distribution of Census-Tract-Level Evacuation Compliance Rates}
    \label{fig:spatial}
\end{figure}

In Figure \ref{fig:spatial}, we presented the spatial distribution of the census-tract-level evacuation compliance rates (computed by using Eqn. (3)). We found that several tracts within/near the southern boundary of the fire perimeter had very high evacuation compliance rates (i.e., above 80\%). This is consistent with the Protective Action Decision Model (PADM) by Lindell and Perry that both environmental cues (close to fire perimeter) and warning messages (evacuation warning/order) have strong influences on people's evacuation decision-making in emergencies \citep{lindell2012protective}. However, we observed that the large tract at the top right corner of the county has an evacuation compliance rate around 50\%, despite its proximity to the fire and being under an evacuation order. Future research into land use of this area and comparison with survey findings from the same fire \citep{erica2021, zhao2021} is needed to explain this result further. Moreover, we observed relatively high compliance of evacuation orders among most tracts in southwest Sonoma County, CA, even though they were not close to the fire perimeter. This result also aligns with local residents' high perception of wildfire risks \citep{erica2021}. 

\subsubsection{Proportion of Different Groups Within Evacuation Warning/Order Zones}

We first computed the overall proportion of different groups within evacuation zones. Note that shadow evacuees are people who chose to evacuate while living outside but near the evacuation zones (within a 5-mile buffer of the zones' boundaries), so they were not included in this analysis. We found that 35\% of the residents evacuated, while 42\% of them stayed in place (and the uncategorized people\footnote{All cases that do not fit the criteria of self-evacuee, shadow evacuee, evacuee under warning, ordered evacuee, and non-evacuee.} accounted for 23\%). In other words, among categorized individuals, 46\% of them evacuated and 54\% did not evacuate. In our questionnaire survey study about the Kincade Fire evacuation process \citep{erica2021}, we found that around 80\% of survey respondents evacuated eventually, which is equivalent to an evacuation compliance rate approximately 34\% higher than the rate inferred from the GPS data (46\%). Note that some survey respondents were not located in the evacuation zones (at the time of the fire). Since both survey data and GPS data have sampling bias issues \citep{erica2021,carto2021}, more research is needed to explain the discrepancy between the two.

\begin{figure}[H]
    \centering
    \includegraphics[width=0.8\textwidth]{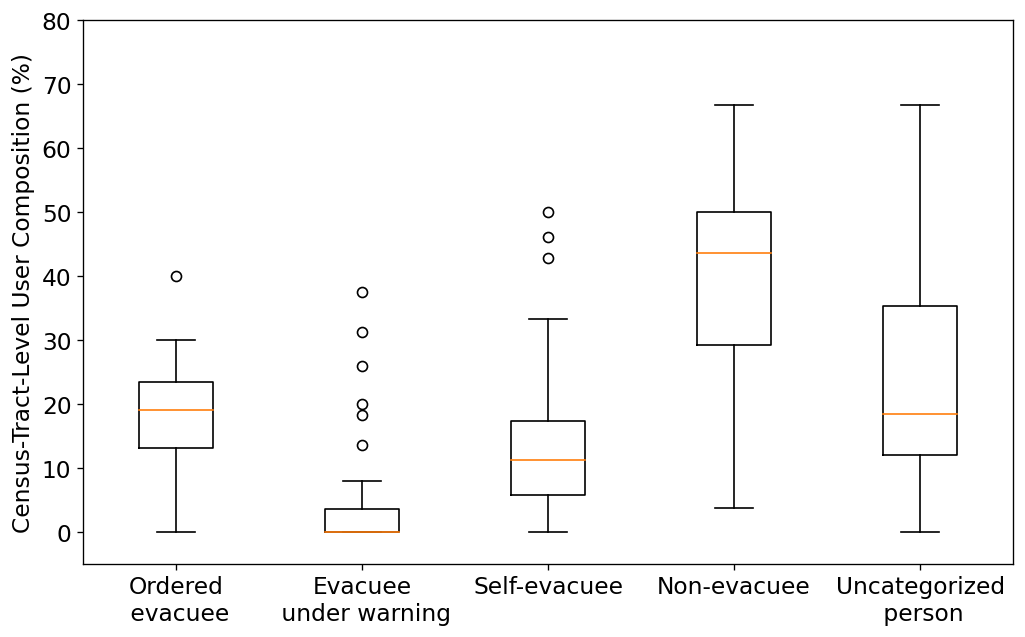}
    \caption{Percentages of Different Groups in the Census Tracts Under Investigation}
    \label{fig:boxplot}
\end{figure}

For each census tract, we computed its resident composition (i.e., percentages of different groups), and then presented the variations of census-tract-level resident composition in Figure \ref{fig:boxplot}. These boxplots show how the percentages of individuals who evacuated or not vary dramatically in the census tracts under investigation. For instance, less than 10\% of individuals did not evacuate in some areas while almost 70\% of individual took the same protective action in other areas. This illustrates the percentage obtained in the questionnaire survey study are within the percentage intervals illustrated in Figure \ref{fig:boxplot}.


\section{Discussion and Conclusion}

By leveraging a large-scale GPS dataset, this study developed a novel methodology to systematically analyze the wildfire evacuation process and identify different groups of evacuees (i.e., self-evacuee, shadow evacuee, evacuee under warning, and ordered evacuee). We tested and demonstrated the proposed methodology with a case study of the 2019 Kincade Fire in Sonoma County, CA. The findings of this study can be used by emergency managers and planners to better understand human behavior in wildfires and thus develop targeted public outreach campaigns, training protocols, and emergency communication strategies to prepare WUI households for future wildfires.

An important finding of this study is that among all groups of evacuees, self-evacuees and shadow evacuees consisted of more than half of evacuees during the Kincade Fire. This result suggests that the local residents were sensitive to wildfire risks, likely because they had prior experience with the 2017 Tubbs Fire (which burned parts of Sonoma, Napa, and Lake counties and was the most destructive wildfire in the history of California until 2017) \citep{erica2021}. This trend is in line with the literature showing that previous wildfire experiences increase both people's risk perception and can increase their probability of householders to evacuate during a natural disaster \citep{benight2004collective,lovreglio2019modelling}. Shadow evacuation indicates people who evacuated even though they might not have been required to \citep{dash2007evacuation}. Shadow evacuation is often considered a problem during hurricane evacuations, since it may lead to traffic congestion and delayed evacuation for people who live in the evacuation zones \citep{zhang2020effects}. This study is the first attempt to borrow this concept in a wildfire evacuation study. 

Furthermore, within the evacuation zones, the total evacuation compliance rate is around 50\%, which shows some discrepancy from the results obtained from a separate survey study for the same fire \citep{zhao2021}; however, it is worth noting that the evacuation compliance rate varies significantly across space. One possible explanation is that many evacuees were classified as uncategorized people, due to the lack of supporting evidence. For example, some evacuees might not use the apps that recorded their locations (so no GPS pins) after they left home for evacuation. Some evacuees returned home early (even before the warning was lifted), but we did not count them as evacuees according to our evacuation-behavior inference algorithm (see Figure \ref{fig:algo_eva}). Additionally, as discussed in \citet{carto2021}, the GPS data may have potential geographic (the sample may overrepresent certain groups of people), demographic (mobile device penetration and usage is not the same in rural versus urban communities \citep{heimerl2015analysis}), temporal (the mobile devices represented in the sample may vary over time), and/or behavioral biases (only certain apps collect location data). On the other hand, the survey data tends to have reporting bias \citep{babbie2020practice}. For example, it is possible the Kincade Fire survey \citep{zhao2021} oversampled householders who decided to evacuate. Additionally, compared to another survey conducted for the 2017 Northern California Wildfires, which included the 2017 Tubbs Fire, \citet{wong2020review} reported approximately 47\% evacuation compliance rate, which shows comparable outcome to the GPS-data-based estimate for the 2019 Kincade Fire\footnote{Both the 2017 Tubbs Fire and the 2019 Kincade Fire significantly impacted Sonoma County, CA.}. Future research is required to better understand the biases of GPS and survey data and to investigate the reasons behind this discrepancy in evacuation compliance rate.

There are some limitations of this study. First, for some census tracts, we have less than 20 inferred residents (or, home locations), making the evacuation analyses of these tracts less reliable. Future studies should investigate why these tracts have small sample sizes and may consider merging multiple GPS datasets from different providers/sources to increase the sample size for analysis. Second, we chose a five-mile buffer around the evacuation zone to analyze shadow evacuation behavior. We also assumed the time threshold for users who lived outside the evacuation zone is equal to two days and the distance threshold to five miles. In future work, these assumptions will be assessed (and adjusted if necessary) by conducting a sensitivity analysis of the key parameters. Third, the GPS data have different types of biases as discussed above, and a new methodology needs to be created to reduce the bias and generate more realistic results for more effective decision-making. Future work can also include developing a model to analyze the relationship between different important factors (e.g., sociodemographics, distance to the fire perimeter, and timing of the evacuation warning/order) and the wildfire evacuation compliance rate, in order to extract key insights for emergency planning and management.

\section*{Acknowledgements}
This work was performed under the following financial assistance award No. 60NANB20D182 from U.S. Department of Commerce, National Institute of Standards and Technology (NIST). Any opinions, findings, conclusions, or recommendations expressed in this material are those of the authors and do not necessarily reflect the views of NIST. The authors would like to thank Dr. Nancy A. Brown, the Emergency Manager of Sonoma County, CA for providing valuable inputs for the study.

\section*{CRediT authorship contribution statement}
\textbf{Xilei Zhao:} Conceptualization; Methodology; Formal analysis; Resources; Data Curation; Writing - Original Draft; Supervision; Project Administration; Funding Acquisition. \textbf{Yiming Xu:} Methodology; Software; Validation; Formal analysis; Investigation; Data Curation; Writing - Original Draft; Visualization. \textbf{Ruggiero Lovreglio:} Conceptualization; Methodology; Writing - Review \& Editing; Funding Acquisition. \textbf{Erica Kuligowski:} Conceptualization; Methodology; Writing - Review \& Editing; Funding Acquisition. \textbf{Daniel Nilsson:} Conceptualization; Methodology; Writing - Review \& Editing; Funding Acquisition. \textbf{Thomas Cova:} Conceptualization; Methodology; Writing - Review \& Editing. \textbf{Alex Wu:} Software; Formal analysis; Investigation; Writing - Original Draft; Visualization. \textbf{Xiang Yan:} Conceptualization; Methodology; Writing - Review \& Editing.

\section*{Declaration of Competing Interest}
None.






\bibliographystyle{abbrvnat}
\biboptions{semicolon,round,sort,authoryear}
\bibliography{sample.bib}

\begin{thebibliography}{63}
\providecommand{\natexlab}[1]{#1}
\providecommand{\url}[1]{\texttt{#1}}
\expandafter\ifx\csname urlstyle\endcsname\relax
  \providecommand{\doi}[1]{doi: #1}\else
  \providecommand{\doi}{doi: \begingroup \urlstyle{rm}\Url}\fi

\bibitem[Ahas et~al.(2010)Ahas, Silm, J{\"a}rv, Saluveer, and
  Tiru]{ahas2010using}
R.~Ahas, S.~Silm, O.~J{\"a}rv, E.~Saluveer, and M.~Tiru.
\newblock Using mobile positioning data to model locations meaningful to users
  of mobile phones.
\newblock \emph{Journal of Urban Technology}, 17\penalty0 (1):\penalty0 3--27,
  2010.

\bibitem[Alexander et~al.(2015)Alexander, Jiang, Murga, and
  Gonz{\'a}lez]{alexander2015origin}
L.~Alexander, S.~Jiang, M.~Murga, and M.~C. Gonz{\'a}lez.
\newblock Origin--destination trips by purpose and time of day inferred from
  mobile phone data.
\newblock \emph{Transportation Research Part C: Emerging Technologies},
  58:\penalty0 240--250, 2015.

\bibitem[Babbie(2020)]{babbie2020practice}
E.~R. Babbie.
\newblock \emph{The practice of social research}.
\newblock Cengage learning, 2020.

\bibitem[Benight(2004)]{benight2004collective}
C.~C. Benight.
\newblock Collective efficacy following a series of natural disasters.
\newblock \emph{Anxiety, Stress \& Coping}, 17\penalty0 (4):\penalty0 401--420,
  2004.

\bibitem[Boustras et~al.(2017)Boustras, Ronchi, and Rein]{boustras2017fires}
G.~Boustras, E.~Ronchi, and G.~Rein.
\newblock Fires: fund research for citizen safety.
\newblock \emph{Nature}, 551\penalty0 (7680):\penalty0 300--301, 2017.

\bibitem[Calabrese et~al.(2013)Calabrese, Diao, Di~Lorenzo, Ferreira~Jr, and
  Ratti]{calabrese2013understanding}
F.~Calabrese, M.~Diao, G.~Di~Lorenzo, J.~Ferreira~Jr, and C.~Ratti.
\newblock Understanding individual mobility patterns from urban sensing data: A
  mobile phone trace example.
\newblock \emph{Transportation Research Part C: Emerging Technologies},
  26:\penalty0 301--313, 2013.

\bibitem[Census~Bureau(2019)]{census2019}
U.~S. Census~Bureau.
\newblock Quickfacts: Sonoma county, california;.
\newblock 2019.
\newblock URL
  \url{https://www.census.gov/quickfacts/fact/table/sonomacountycalifornia,CA/PST045219}.

\bibitem[Chen et~al.(2016)Chen, Ma, Susilo, Liu, and Wang]{chen2016promises}
C.~Chen, J.~Ma, Y.~Susilo, Y.~Liu, and M.~Wang.
\newblock The promises of big data and small data for travel behavior (aka
  human mobility) analysis.
\newblock \emph{Transportation Research Part C: Emerging Technologies},
  68:\penalty0 285--299, 2016.

\bibitem[Cova et~al.(2011)Cova, Dennison, and Drews]{cova2011modeling}
T.~J. Cova, P.~E. Dennison, and F.~A. Drews.
\newblock Modeling evacuate versus shelter-in-place decisions in wildfires.
\newblock \emph{Sustainability}, 3\penalty0 (10):\penalty0 1662--1687, 2011.

\bibitem[Dash and Gladwin(2007)]{dash2007evacuation}
N.~Dash and H.~Gladwin.
\newblock Evacuation decision making and behavioral responses: Individual and
  household.
\newblock \emph{Natural Hazards Review}, 8\penalty0 (3):\penalty0 69--77, 2007.

\bibitem[Demissie et~al.(2019)Demissie, Phithakkitnukoon, Kattan, and
  Farhan]{demissie2019understanding}
M.~G. Demissie, S.~Phithakkitnukoon, L.~Kattan, and A.~Farhan.
\newblock Understanding human mobility patterns in a developing country using
  mobile phone data.
\newblock \emph{Data Science Journal}, 18\penalty0 (1), 2019.

\bibitem[Dennison et~al.(2007)Dennison, Cova, and Mortiz]{dennison2007wuivac}
P.~E. Dennison, T.~J. Cova, and M.~A. Mortiz.
\newblock Wuivac: a wildland-urban interface evacuation trigger model applied
  in strategic wildfire scenarios.
\newblock \emph{Natural Hazards}, 41\penalty0 (1):\penalty0 181--199, 2007.

\bibitem[Dixit et~al.(2011)Dixit, Montz, and Wolshon]{dixit2011validation}
V.~Dixit, T.~Montz, and B.~Wolshon.
\newblock Validation techniques for region-level microscopic mass evacuation
  traffic simulations.
\newblock \emph{Transportation research record}, 2229\penalty0 (1):\penalty0
  66--74, 2011.

\bibitem[Freedman(2020)]{westernfires}
A.~Freedman.
\newblock Western wildfires: An 'unprecedented' climate change fueled event,
  experts say.
\newblock \emph{The Washington Post}, 2020.

\bibitem[Fu(2004)]{fu2004development}
H.~Fu.
\newblock \emph{Development of dynamic travel demand models for hurricane
  evacuation}.
\newblock Louisiana State University and Agricultural \& Mechanical College,
  2004.

\bibitem[Fu et~al.(2007)Fu, Wilmot, Zhang, and Baker]{fu2007modeling}
H.~Fu, C.~G. Wilmot, H.~Zhang, and E.~J. Baker.
\newblock Modeling the hurricane evacuation response curve.
\newblock \emph{Transportation Research Record}, 2022\penalty0 (1):\penalty0
  94--102, 2007.

\bibitem[Gladwin and Peacock(1997)]{Gladwin1997}
H.~Gladwin and W.~Peacock.
\newblock {Warning and Evacuation: A Night for Hard Houses}.
\newblock \emph{Hurricane Andrew: Ethnicity, gender and the sociology of
  disasters}, pages 52--74, 1997.

\bibitem[Grajdura et~al.(2021)Grajdura, Qian, and
  Niemeier]{grajdura2021awareness}
S.~Grajdura, X.~Qian, and D.~Niemeier.
\newblock Awareness, departure, and preparation time in no-notice wildfire
  evacuations.
\newblock \emph{Safety Science}, 139:\penalty0 105258, 2021.

\bibitem[Hayano and Adachi(2013)]{hayano2013estimation}
R.~S. Hayano and R.~Adachi.
\newblock Estimation of the total population moving into and out of the 20 km
  evacuation zone during the fukushima npp accident as calculated using
  “auto-gps” mobile phone data.
\newblock \emph{Proceedings of the Japan Academy, Series B}, 89\penalty0
  (5):\penalty0 196--199, 2013.

\bibitem[Heimerl et~al.(2015)Heimerl, Menon, Hasan, Ali, Brewer, and
  Parikh]{heimerl2015analysis}
K.~Heimerl, A.~Menon, S.~Hasan, K.~Ali, E.~Brewer, and T.~Parikh.
\newblock Analysis of smartphone adoption and usage in a rural community
  cellular network.
\newblock In \emph{Proceedings of the Seventh International Conference on
  Information and Communication Technologies and Development}, pages 1--4,
  2015.

\bibitem[Horanont et~al.(2013)Horanont, Witayangkurn, Sekimoto, and
  Shibasaki]{horanont2013large}
T.~Horanont, A.~Witayangkurn, Y.~Sekimoto, and R.~Shibasaki.
\newblock Large-scale auto-gps analysis for discerning behavior change during
  crisis.
\newblock \emph{IEEE Intelligent Systems}, 28\penalty0 (4):\penalty0 26--34,
  2013.

\bibitem[Kuligowski(2021)]{kuligowski2021evacuation}
E.~Kuligowski.
\newblock Evacuation decision-making and behavior in wildfires: Past research,
  current challenges and a future research agenda.
\newblock \emph{Fire Safety Journal}, 120:\penalty0 103129, 2021.

\bibitem[Kuligowski et~al.(2020)Kuligowski, Walpole, Lovreglio, and
  McCaffrey]{kuligowski2020modelling}
E.~D. Kuligowski, E.~H. Walpole, R.~Lovreglio, and S.~McCaffrey.
\newblock Modelling evacuation decision-making in the 2016 chimney tops 2 fire
  in gatlinburg, tn.
\newblock \emph{International Journal of Wildland Fire}, 29\penalty0
  (12):\penalty0 1120--1132, 2020.

\bibitem[Kuligowski et~al.(Under Review)Kuligowski, Zhao, Lovreglio, Xu, Yang,
  Westbury, Nilsson, and Brown]{erica2021}
E.~D. Kuligowski, X.~Zhao, R.~Lovreglio, N.~Xu, K.~Yang, A.~Westbury,
  D.~Nilsson, and N.~Brown.
\newblock Modeling evacuation decision-making in the 2019 kincade fire in
  california.
\newblock \emph{Safety Science}, Under Review.

\bibitem[Li et~al.(2014)Li, Cheng, Duan, Yang, and Guo]{li2014framework}
W.~Li, X.~Cheng, Z.~Duan, D.~Yang, and G.~Guo.
\newblock A framework for spatial interaction analysis based on large-scale
  mobile phone data.
\newblock \emph{Computational Intelligence and Neuroscience}, 2014, 2014.

\bibitem[Lindell and Perry(2012)]{lindell2012protective}
M.~K. Lindell and R.~W. Perry.
\newblock The protective action decision model: theoretical modifications and
  additional evidence.
\newblock \emph{Risk Analysis: An International Journal}, 32\penalty0
  (4):\penalty0 616--632, 2012.

\bibitem[Liu et~al.(2010)Liu, Stanturf, and Goodrick]{liu2010trends}
Y.~Liu, J.~Stanturf, and S.~Goodrick.
\newblock Trends in global wildfire potential in a changing climate.
\newblock \emph{Forest Ecology and Management}, 259\penalty0 (4):\penalty0
  685--697, 2010.

\bibitem[Lovreglio et~al.(2019)Lovreglio, Kuligowski, Gwynne, and
  Strahan]{lovreglio2019modelling}
R.~Lovreglio, E.~Kuligowski, S.~Gwynne, and K.~Strahan.
\newblock A modelling framework for householder decision-making for wildfire
  emergencies.
\newblock \emph{International Journal of Disaster Risk Reduction}, 41:\penalty0
  101274, 2019.

\bibitem[Lovreglio et~al.(2020)Lovreglio, Kuligowski, Walpole, Link, and
  Gwynne]{lovreglio2020calibrating}
R.~Lovreglio, E.~Kuligowski, E.~Walpole, E.~Link, and S.~Gwynne.
\newblock Calibrating the wildfire decision model using hybrid choice
  modelling.
\newblock \emph{International Journal of Disaster Risk Reduction}, 50:\penalty0
  101770, 2020.

\bibitem[McCaffrey et~al.(2018)McCaffrey, Wilson, and
  Konar]{mccaffrey2018should}
S.~McCaffrey, R.~Wilson, and A.~Konar.
\newblock Should i stay or should i go now? or should i wait and see?
  influences on wildfire evacuation decisions.
\newblock \emph{Risk Analysis}, 38\penalty0 (7):\penalty0 1390--1404, 2018.

\bibitem[McLennan et~al.(2019)McLennan, Ryan, Bearman, and
  Toh]{mclennan2019should}
J.~McLennan, B.~Ryan, C.~Bearman, and K.~Toh.
\newblock Should we leave now? behavioral factors in evacuation under wildfire
  threat.
\newblock \emph{Fire technology}, 55\penalty0 (2):\penalty0 487--516, 2019.

\bibitem[Mitchell et~al.(2007)Mitchell, Cutter, Edmonds,
  et~al.]{mitchell2007improving}
J.~T. Mitchell, S.~L. Cutter, A.~S. Edmonds, et~al.
\newblock Improving shadow evacuation management: Case study of the
  graniteville, south carolina, chlorine spill.
\newblock \emph{Journal of Emergency Management}, 5\penalty0 (1):\penalty0
  28--34, 2007.

\bibitem[Murray-Tuite and Wolshon(2013)]{murray2013evacuation}
P.~Murray-Tuite and B.~Wolshon.
\newblock Evacuation transportation modeling: An overview of research,
  development, and practice.
\newblock \emph{Transportation Research Part C: Emerging Technologies},
  27:\penalty0 25--45, 2013.

\bibitem[Ozbay et~al.(2012)Ozbay, Yazici, Iyer, Li, Ozguven, and
  Carnegie]{ozbay2012use}
K.~Ozbay, M.~A. Yazici, S.~Iyer, J.~Li, E.~E. Ozguven, and J.~A. Carnegie.
\newblock Use of regional transportation planning tool for modeling emergency
  evacuation: Case study of northern new jersey.
\newblock \emph{Transportation research record}, 2312\penalty0 (1):\penalty0
  89--97, 2012.

\bibitem[Quddus et~al.(2007)Quddus, Ochieng, and Noland]{quddus2007current}
M.~A. Quddus, W.~Y. Ochieng, and R.~B. Noland.
\newblock Current map-matching algorithms for transport applications:
  State-of-the art and future research directions.
\newblock \emph{Transportation Research Part C: Emerging Technologies},
  15\penalty0 (5):\penalty0 312--328, 2007.

\bibitem[Radeloff et~al.(2018)Radeloff, Helmers, Kramer, Mockrin, Alexandre,
  Bar-Massada, Butsic, Hawbaker, Martinuzzi, Syphard,
  et~al.]{radeloff2018rapid}
V.~C. Radeloff, D.~P. Helmers, H.~A. Kramer, M.~H. Mockrin, P.~M. Alexandre,
  A.~Bar-Massada, V.~Butsic, T.~J. Hawbaker, S.~Martinuzzi, A.~D. Syphard,
  et~al.
\newblock Rapid growth of the us wildland-urban interface raises wildfire risk.
\newblock \emph{Proceedings of the National Academy of Sciences}, 115\penalty0
  (13):\penalty0 3314--3319, 2018.

\bibitem[Ronchi et~al.(2019)Ronchi, Gwynne, Rein, Intini, and
  Wadhwani]{ronchi2019open}
E.~Ronchi, S.~M. Gwynne, G.~Rein, P.~Intini, and R.~Wadhwani.
\newblock An open multi-physics framework for modelling wildland-urban
  interface fire evacuations.
\newblock \emph{Safety Science}, 118:\penalty0 868--880, 2019.

\bibitem[Song et~al.(2013)Song, Zhang, Sekimoto, Horanont, Ueyama, and
  Shibasaki]{song2013intelligent}
X.~Song, Q.~Zhang, Y.~Sekimoto, T.~Horanont, S.~Ueyama, and R.~Shibasaki.
\newblock Intelligent system for human behavior analysis and reasoning
  following large-scale disasters.
\newblock \emph{IEEE Intelligent Systems}, 28\penalty0 (4):\penalty0 35--42,
  2013.

\bibitem[{Sonoma Operational Area and the County of Sonoma, Department of
  Emergency Management}(2020)]{sonoma2020report}
{Sonoma Operational Area and the County of Sonoma, Department of Emergency
  Management}.
\newblock 2019 kincade fire after action report.
\newblock \emph{2019 Kincade Fire After-Action Report and Improvement Plan},
  2020.

\bibitem[Strahan and Watson(2019)]{strahan2019protective}
K.~Strahan and S.~J. Watson.
\newblock The protective action decision model: When householders choose their
  protective response to wildfire.
\newblock \emph{Journal of Risk Research}, 22\penalty0 (12):\penalty0
  1602--1623, 2019.

\bibitem[Tettamanti et~al.(2012)Tettamanti, Demeter, and
  Varga]{tettamanti2012route}
T.~Tettamanti, H.~Demeter, and I.~Varga.
\newblock Route choice estimation based on cellular signaling data.
\newblock \emph{Acta Polytechnica Hungarica}, 9\penalty0 (4):\penalty0
  207--220, 2012.

\bibitem[Toledo et~al.(2018)Toledo, Marom, Grimberg, and
  Bekhor]{toledo2018analysis}
T.~Toledo, I.~Marom, E.~Grimberg, and S.~Bekhor.
\newblock Analysis of evacuation behavior in a wildfire event.
\newblock \emph{International journal of disaster risk reduction}, 31:\penalty0
  1366--1373, 2018.

\bibitem[Trufero and Koschinsky(2021)]{carto2021}
J.~P. Trufero and J.~Koschinsky.
\newblock Making human mobility models fair, inclusive, \& private.
\newblock \emph{CARTO}, 2021.
\newblock URL
  \url{https://carto.com/blog/making-human-mobility-models-fair-inclusive-private/}.

\bibitem[Vaiciulyte et~al.(2021)Vaiciulyte, Hulse, Veeraswamy, and
  Galea]{vaiciulyte2021cross}
S.~Vaiciulyte, L.~M. Hulse, A.~Veeraswamy, and E.~R. Galea.
\newblock Cross-cultural comparison of behavioural itinerary actions and times
  in wildfire evacuations.
\newblock \emph{Safety Science}, 135:\penalty0 105122, 2021.

\bibitem[Vanhoof et~al.(2018)Vanhoof, Reis, Ploetz, and
  Smoreda]{vanhoof2018assessing}
M.~Vanhoof, F.~Reis, T.~Ploetz, and Z.~Smoreda.
\newblock Assessing the quality of home detection from mobile phone data for
  official statistics.
\newblock \emph{arXiv preprint arXiv:1809.07567}, 2018.

\bibitem[Wang et~al.(2010)Wang, Calabrese, Di~Lorenzo, and
  Ratti]{wang2010transportation}
H.~Wang, F.~Calabrese, G.~Di~Lorenzo, and C.~Ratti.
\newblock Transportation mode inference from anonymized and aggregated mobile
  phone call detail records.
\newblock In \emph{13th International IEEE Conference on Intelligent
  Transportation Systems}, pages 318--323. IEEE, 2010.

\bibitem[Wang et~al.(2018)Wang, He, and Leung]{wang2018applying}
Z.~Wang, S.~Y. He, and Y.~Leung.
\newblock Applying mobile phone data to travel behaviour research: A literature
  review.
\newblock \emph{Travel Behaviour and Society}, 11:\penalty0 141--155, 2018.

\bibitem[Wolshon and Marchive~III(2007)]{wolshon2007emergency}
B.~Wolshon and E.~Marchive~III.
\newblock Emergency planning in the urban-wildland interface: Subdivision-level
  analysis of wildfire evacuations.
\newblock \emph{Journal of Urban Planning and Development}, 133\penalty0
  (1):\penalty0 73--81, 2007.

\bibitem[Wong et~al.(2020{\natexlab{a}})Wong, Broader, and
  Shaheen]{wong2020review}
S.~D. Wong, J.~C. Broader, and S.~A. Shaheen.
\newblock Review of california wildfire evacuations from 2017 to 2019.
\newblock 2020{\natexlab{a}}.

\bibitem[Wong et~al.(2020{\natexlab{b}})Wong, Broader, Walker, and
  Shaheen]{wong2020understanding}
S.~D. Wong, J.~C. Broader, J.~L. Walker, and S.~A. Shaheen.
\newblock Understanding california wildfire evacuee behavior and joint
  choice-making.
\newblock 2020{\natexlab{b}}.

\bibitem[Woo et~al.(2017)Woo, Hui, Ren, Gan, and Kim]{woo2017reconstructing}
M.~Woo, K.~T.~Y. Hui, K.~Ren, K.~E. Gan, and A.~Kim.
\newblock Reconstructing an emergency evacuation by ground and air the wildfire
  in fort mcmurray, alberta, canada.
\newblock \emph{Transportation Research Record}, 2604\penalty0 (1):\penalty0
  63--70, 2017.

\bibitem[Xu et~al.(2015)Xu, Shaw, Zhao, Yin, Fang, and Li]{xu2015understanding}
Y.~Xu, S.-L. Shaw, Z.~Zhao, L.~Yin, Z.~Fang, and Q.~Li.
\newblock Understanding aggregate human mobility patterns using passive mobile
  phone location data: a home-based approach.
\newblock \emph{Transportation}, 42\penalty0 (4):\penalty0 625--646, 2015.

\bibitem[Yabe and Ukkusuri(2020)]{yabe2020effects}
T.~Yabe and S.~V. Ukkusuri.
\newblock Effects of income inequality on evacuation, reentry and segregation
  after disasters.
\newblock \emph{Transportation Research Part D: Transport and Environment},
  82:\penalty0 102260, 2020.

\bibitem[Yabe et~al.(2016)Yabe, Tsubouchi, Sudo, and
  Sekimoto]{yabe2016estimating}
T.~Yabe, K.~Tsubouchi, A.~Sudo, and Y.~Sekimoto.
\newblock Estimating evacuation hotspots using gps data: What happened after
  the large earthquakes in kumamoto, japan.
\newblock In \emph{Proc. of the 5th International Workshop on Urban Computing},
  volume~81, pages 1--5, 2016.

\bibitem[Yabe et~al.(2019)Yabe, Sekimoto, Tsubouchi, and
  Ikemoto]{yabe2019cross}
T.~Yabe, Y.~Sekimoto, K.~Tsubouchi, and S.~Ikemoto.
\newblock Cross-comparative analysis of evacuation behavior after earthquakes
  using mobile phone data.
\newblock \emph{PLoS one}, 14\penalty0 (2):\penalty0 e0211375, 2019.

\bibitem[Yazici and Ozbay(2008)]{yazici2008evacuation}
M.~A. Yazici and K.~Ozbay.
\newblock Evacuation modelling in the united states: Does the demand model
  choice matter?
\newblock \emph{Transport Reviews}, 28\penalty0 (6):\penalty0 757--779, 2008.

\bibitem[Yu et~al.(2020)Yu, Li, Yang, and Zhang]{yu2020mobile}
Q.~Yu, W.~Li, D.~Yang, and H.~Zhang.
\newblock Mobile phone data in urban commuting: A network community
  detection-based framework to unveil the spatial structure of commuting
  demand.
\newblock \emph{Journal of Advanced Transportation}, 2020, 2020.

\bibitem[Zeigler et~al.(1981)Zeigler, Brunn, and
  Johnson~Jr]{zeigler1981evacuation}
D.~J. Zeigler, S.~D. Brunn, and J.~H. Johnson~Jr.
\newblock Evacuation from a nuclear technological disaster.
\newblock \emph{Geographical review}, pages 1--16, 1981.

\bibitem[Zhang et~al.(2016)Zhang, Manjourides, Cohen, Hu, and
  Jiang]{zhang2016spatial}
Z.~Zhang, J.~Manjourides, T.~Cohen, Y.~Hu, and Q.~Jiang.
\newblock Spatial measurement errors in the field of spatial epidemiology.
\newblock \emph{International Journal of Health Geographics}, 15\penalty0
  (1):\penalty0 1--12, 2016.

\bibitem[Zhang et~al.(2020)Zhang, Herrera, Tuncer, Parr, Shapouri, and
  Wolshon]{zhang2020effects}
Z.~Zhang, N.~Herrera, E.~Tuncer, S.~Parr, M.~Shapouri, and B.~Wolshon.
\newblock Effects of shadow evacuation on megaregion evacuations.
\newblock \emph{Transportation Research Part D: Transport and Environment},
  83:\penalty0 102295, 2020.

\bibitem[Zhao et~al.(2020)Zhao, Liu, Yu, and Hu]{zhao2020long}
P.~Zhao, D.~Liu, Z.~Yu, and H.~Hu.
\newblock Long commutes and transport inequity in china’s growing megacity:
  new evidence from beijing using mobile phone data.
\newblock \emph{Travel Behaviour and Society}, 20:\penalty0 248--263, 2020.

\bibitem[Zhao et~al.(2021{\natexlab{a}})Zhao, Lovreglio, Kuligowski, and
  Nilsson]{zhao2021using}
X.~Zhao, R.~Lovreglio, E.~Kuligowski, and D.~Nilsson.
\newblock Using artificial intelligence for safe and effective wildfire
  evacuations.
\newblock \emph{Fire Technology}, 57\penalty0 (2):\penalty0 483--485,
  2021{\natexlab{a}}.

\bibitem[Zhao et~al.(2021{\natexlab{b}})Zhao, Xu, Yang, Kuligowski, Lovreglio,
  Nilsson, and Brown]{zhao2021}
X.~Zhao, N.~Xu, K.~Yang, E.~D. Kuligowski, R.~Lovreglio, D.~Nilsson, and N.~A.
  Brown.
\newblock Modeling evacuation behavior in the 2019 kincade fire, sonoma county,
  california.
\newblock \emph{Natural Hazards Center Quick Response Grant Report Series},
  326, 2021{\natexlab{b}}.

\end{thebibliography}







\end{document}